\documentclass[reprint,amsmath,amssymb,aps,pre]{revtex4-1}
\usepackage{graphicx} 
\usepackage{dcolumn}   
\usepackage{bm}        
\usepackage{amssymb}   
\usepackage{mathtools}
\usepackage{amsmath}
\usepackage{color}
\usepackage{dblfloatfix} 
\usepackage{hyperref}
\hypersetup{
    citecolor={cyan},
    colorlinks=true,
    linkcolor=cyan,
    filecolor=magenta,      
    urlcolor=cyan,
    pdfpagemode=FullScreen,
    }
\begin{document}

\title{ A slime mold inspired local adaptive mechanism for flow networks}
\author{Vidyesh Rao Anisetti,$^1$, Ananth Kandala,$^{2}$ J. M. Schwarz$^{1,3}$}

\affiliation{$^1$ Physics Department, Syracuse University, Syracuse, NY 13244 USA\\
$^2$Department of Physics, University of Florida,  FL 32611-8440, USA\\
$^3$ Indian Creek Farm, Ithaca, NY 14850 USA }
\date{\today}
\begin{abstract}
    In the realm of biological flow networks, the ability to dynamically adjust to varying demands is paramount. Drawing inspiration from the remarkable adaptability of Physarum polycephalum, we present a novel physical mechanism tailored to optimize flow networks. Central to our approach is the principle that each network component—specifically, the tubes— harnesses locally available information to collectively minimize a global cost function. Our findings underscore the scalability of this mechanism, making it feasible for larger, more complex networks. We construct a comprehensive phase diagram, pinpointing the specific network parameters under which successful adaptation, or tuning, is realized. There exists a phase boundary in the phase diagram, revealing a distinct satisfiability-unsatisfiability (SAT-UNSAT) phase transition delineating successful and unsuccessful adaptation. 
\end{abstract}

\maketitle

\section{Introduction}
Biological circulatory systems, despite their intricacy, exhibit remarkable adaptability. They adeptly modify various attributes—such as vessel diameter, wall thickness, and the count of micro-vessels cater to the evolving metabolic needs of tissues \cite{hu2012blood,Humphrey2021}. This adaptability can be perceived as an outcome of an optimization mechanism, where a global cost function is optimized. Intriguingly, this optimization is not typically directed by a central authority, but emerges from local interactions between different aspects of the system. For example, prior research illustrates how vascular flow efficiency \cite{ronellenfitsch2016global} can be enhanced by altering attributes, e.g. tube thickness, based on local data such as flow within a tube.\\ \\
Given the circulatory system example, as well as others~\cite{Tero2010,Eytan2003}, a fundamental inquiry centers on how local interaction rules give rise to adaptive behaviors, and how these behaviors subsequently manifest as optimization algorithms in biological systems~\cite{Holland1992}. In this manuscript, we introduce a straightforward physical mechanism that potentially allows biological systems to implement such optimization strategies. Moreover, limitations on such optimization algorithms do indeed exist. For instance, one type of optimization algorithm for a large swath of some parameter space may not be feasible. Here, we explore the limitations on our physical mechanism for adaptation in the context of what is known as a Satisfiability/SAT-Unsatisfiability/UNSAT phase transition~\cite{Mitchell1992,Kirkpatrick1994,Monasson1999}.  A similar transition was found in the study of limits on the multifunctionality of tunable flow and mechanical networks~\cite{Rocks2019}.\\ \\
The specific question we are interested in is: How does one tune the node pressures of a flow network, by altering its tube thickness, via a physical mechanism that uses only local information? To answer this question, we take inspiration from the adaptive behaviour observed in Physarum polycephalum, or slime mold. This organism, in spite of not having a brain or nervous system, is capable of optimizing its network structure to determine the most efficient path between food sources \cite{Tero2010,Tero2008}. Prior research \cite{Alim2017} indicates that slime mold possibly utilizes a form of chemical signaling for this purpose. Upon encountering food, it releases a signaling chemical at the location, which is then propagated throughout the organism via advection. This chemical, in turn, induces an expansion in the tubes through which it is flowing. As a result, the tubes that connect food sources through the shortest path undergo more pronounced expansion compared to those on longer routes, leading to an efficient connection between food sources. This behavior exemplifies how biological systems can harness physical processes to optimize for a specific task such as finding food. 
\begin{figure}[h]
    \centering
    \includegraphics[width=8.5cm]{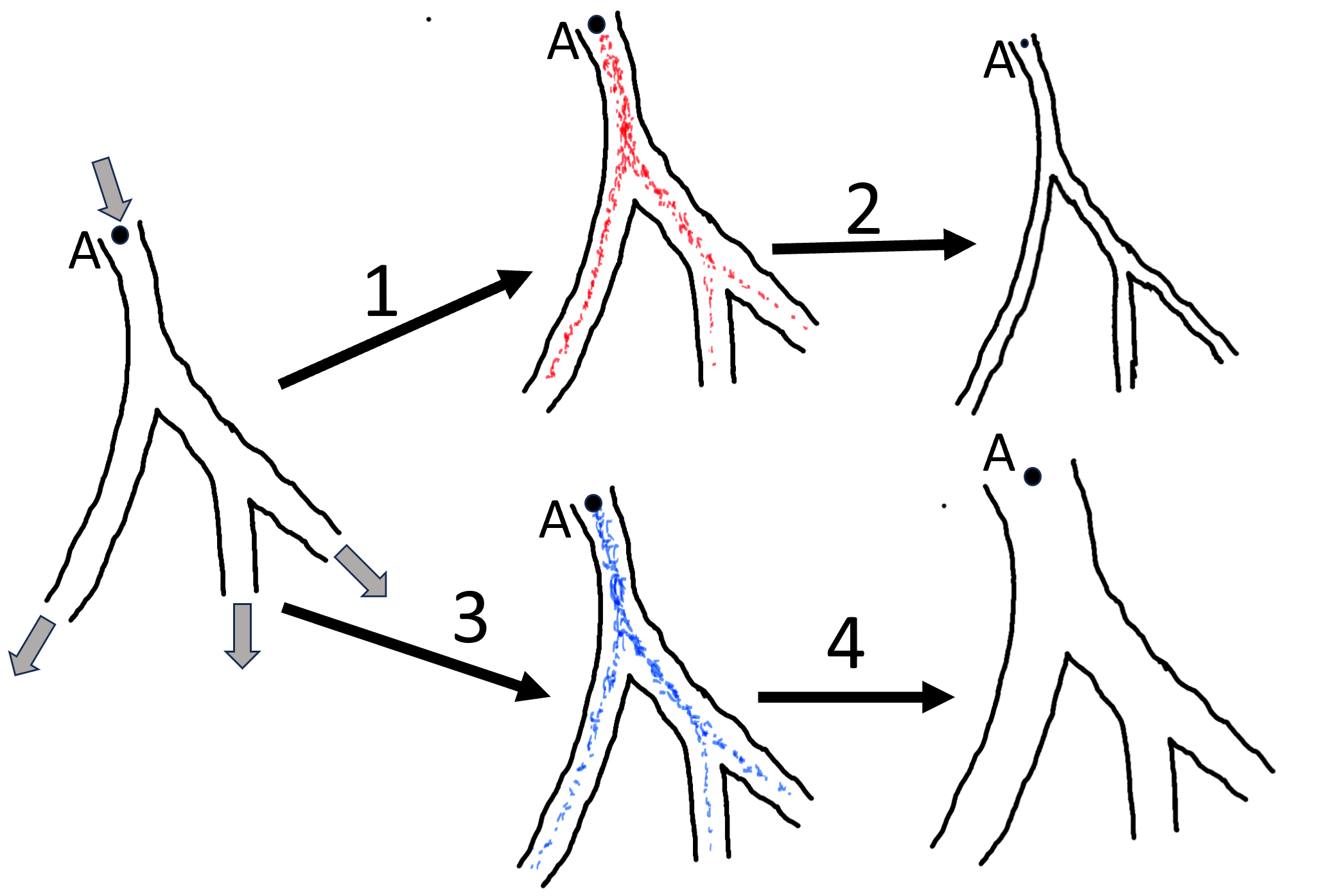}
    \caption{{\it Schematic for an adaptative flow mechanism.} [Arrow 1] To increase the pressure at node A, a specialized chemical (depicted in red) is introduced. This chemical is advectively transported by the fluid flow within the network. The fluid flow direction is indicated by the grey arrows. [Arrow 2] As this chemical traverses the tubes, it interacts with the tube's structure, causing it to constrict and thereby increasing the flow resistance. This results in an increase in pressure at node A. [Arrows 3 \& 4] Conversely, to decrease the pressure at node A, a different chemical is released that dilates the tubes. This dilation reduces flow resistance, leading to a decrease in pressure at node A.}   
    \label{fig:mechanism}
\end{figure}\\ \\
Our system is a flow network - a set of nodes connected by tubes. Fluid flows in and out of the network through some of those nodes. This flow creates pressure values at each node, depending upon the architecture of the network and the conductance of the pipes. The task here is to modify the conductance of these pipes such that the pressure values at some `output' nodes matches the desired pressures.\\ \\
The first thing that we observe is that pressure at a node depends on the resistance to flow downstream. Therefore, to alter the pressure at a node, we must change the conductance of pipes downstream (Fig.~\ref{fig:mechanism}). Consider an output node where we want the pressure to decrease. To do so a chemical at that node gets released which gets carried by the fluid flow. This chemical interacts with the tube such that, when it is flowing through a tube, it increases the conductance of the pipe by making it thicker. This increase in conductance decreases the resistance to flow, which in turn decreases the pressure at the output node. Similarly, when we wish to increase the output node pressure, we must release a different kind of chemical which decreases the conductance of the pipes by making it thinner. Through this mechanism, the entire network undergoes adjustments. Each tube, relying on the locally available information—namely, the chemical concentration within it—fine-tunes the pressures at the output nodes to align with the target values. This localized adjustment is facilitated by our introduced method, where the discrepancy at the output nodes is conveyed into the network through a chemical signal, subsequently influencing the network structure.\\ \\

In what follows, we have assessed the performance of our tuning mechanism across a range of network sizes. The scaling relations observed suggest that our tuning approach remains effective even as the network size increases. Notably, our results suggest a SAT-UNSAT phase transition~\cite{monasson2007introduction} with a discontinuous jump in the fraction of networks that can be successfully adapted at the transition point along with universal scaling features. 
\section{The Tuning Process}
\subsection{The System}
We create networks consisting of N nodes and M edges. We first create a Barabási-Albert network with connection parameter 1 using the python package, NetworkX. This creates a minimally connected network with N nodes and N-1 edges. To create a network with M edges we add M-(N+1) unique edges. \\ \\
We select $j$ boundary nodes, denoted as ${q_1,q_2,...,q_j}$, and apply an external potential $ \mathbf{u}=[u(q_1),u(q_2),...,u(q_j)]^T$. The resulting response potentials at the remaining nodes, termed interior nodes, are calculated by solving the discrete Laplace equation using the Schur compliment method~\cite{Curtis}. From these interior nodes, we identify $k$ output nodes $ { p_1,p_2,..,p_k } $. The potentials at these nodes are represented as ${v(p_1),v(p_2),...,v(p_k)}$. The objective is to adjust this network by altering the conductance of its pipes, aiming to align the output potential vector $\mathbf{v} = [v(p_1),v(p_2),...,v(p_k)]^T$ with a target output potential vector $ \mathbf{v_{des}} = [v_{des}(p_1),v_{des}(p_2),v_{des}(p_3),..,v_{des}(p_k)]^T $. For context, the target output potential vector $ \mathbf{v_{des}}$ is derived by introducing a perturbation to the potential at each node, with the perturbation value sourced from a uniform distribution spanning $[-\Delta,+\Delta]$.
\subsection{Implementation of the Tuning Process :}
\begin{enumerate}
    \item The input potential vector $\mathbf{u}$ is applied at boundary nodes. A supervisor checks the output potentials $\mathbf{v}$ at output nodes and compares them to the vector of desired output potentials $\mathbf{{v}_{des}}$ . 
    \item There are two kinds of chemicals, $s_{+}$ and $s_{-}$. $s_{+}$ increases the conductance of the pipe when it is passing through it, and vice versa for $s_{-}$. We assume that the output nodes release a chemical whose amount is proportional to the difference between the present output potentials and desired output potentials. At $t=0$ for some $a \in O, ~ v(a)  \neq {v}_{des}(a) $, then \begin{align}
        v(a) > {v}_{des}(a) & \Rightarrow s_{+}(a) = \alpha (v(a) - {v}_{des}(a)) \\
        v(a) < {v}_{des}(a) & \Rightarrow s_{-}(a) = \alpha ({v}_{des}(a) - v(a))
    \end{align}
    Where $\alpha$ is the factor that controls the chemical response given by the node to the difference in potentials. Moreover, $s_{+}(a)$ denotes amount of chemical (e.g. no of molecules) at node "a", and $\mathbf{s_{+}}(t)$ denotes the array of chemical amount  at each node at time $t$ .
    \item This chemical is carried by the current in the network. Therefore, in the next time step the chemical flows to the neighbouring nodes of $a$ that are downstream to $a$\footnote{For simplicity we assume that the chemical has negligible diffusion and the only way it can spread is via the network currents}. We call all such downstream neighbours of $a$ as $\mathcal{D}(a)$. Then for all $b \in \mathcal{D}(a): $\begin{align}
        & s_{+}(b,t+1) = s_{+}(a,t) \times \dfrac{i(b,a)}{  \sum_{x \in \mathcal{D}(a) }i(x,a) } \\
        & + {\rm (incoming~chemical~from~other~ nodes)}
    \end{align}
    where $i(x,a)$ represents the current from $a$ to $x$. This is how the entire array $\mathbf{s_{+}}$ and $\mathbf{s_{-}}$ is modified at each time step. Note that all the chemical initially present at $a$ flows downstream after one time step.
    \item Using the above equation, an N $\times$ N array $\hat{S}_{+}$ is generated, where each entry $i,j$ denotes the amount of chemical passing through the pipe $\{i,j\}$ at step $t$ $\xrightarrow{}$ $t+1$. Let $\hat{W}$ denote the conductance matrix of the graph, where each entry $\{i,j\}$ denotes the conductance of that pipe. Then \begin{align}
        & \hat{W}(t+1) = \hat{W}(t) + \beta(\hat{S}_{+} - \hat{S}_{-}),
    \end{align}
    $\beta$ controls the response of the pipe to the passing chemical.
    \item The new potentials are calculated on the interior vertices using $\hat{W}(t+1)$. Again, the supervisor checks if $v(a) = v_{des}(a)$. The chemical takes some time to reach the boundary nodes, where it drains out of the network~\footnote{The current flows into and out of the network through boundary nodes}. Therefore, the total change in potential due to the chemical released at the output nodes is observed after some amount of time. Therefore, we introduce a time delay $\tau$ before releasing the chemical once again~\footnote{This helps the potentials at the output node to converge nicely at the desired potentials. If this delay is not introduced, the output potential oscillates about the desired potential. Also, this helps in avoiding the buildup of excess chemical in the network}.
    \item This process is repeated iteratively.
\end{enumerate} 
\section{Results}
\begin{figure}[h]
    \centering
    \includegraphics[width=8.5cm]{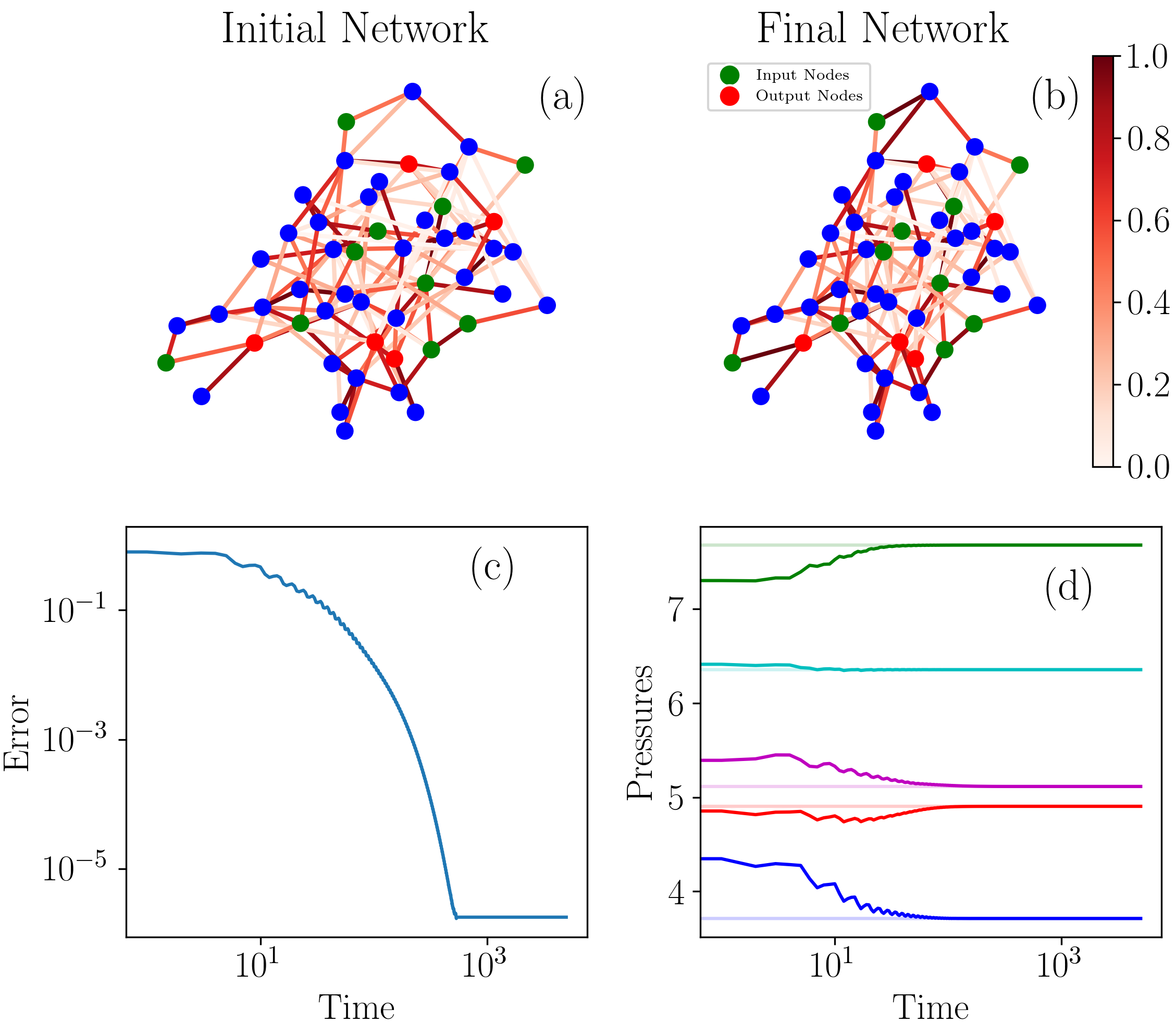}
    \caption{{\it The tuning process.} [a,b] The network undergoes modifications due to the tuning process. The color bar illustrates the conductance values both pre and post-training. [c] A plot of error versus time. The error plateaus since tuning a particular node `$a$' is stopped when $ |v(a)-v_{des}(a)| < 10^{-6}$). [d] Pressures at the output nodes converge to their target values, represented by similarly colored translucent horizontal lines. 
    }   
    \label{fig:1}
\end{figure}

\begin{figure}[h]
    \centering
    \includegraphics[width=8cm]{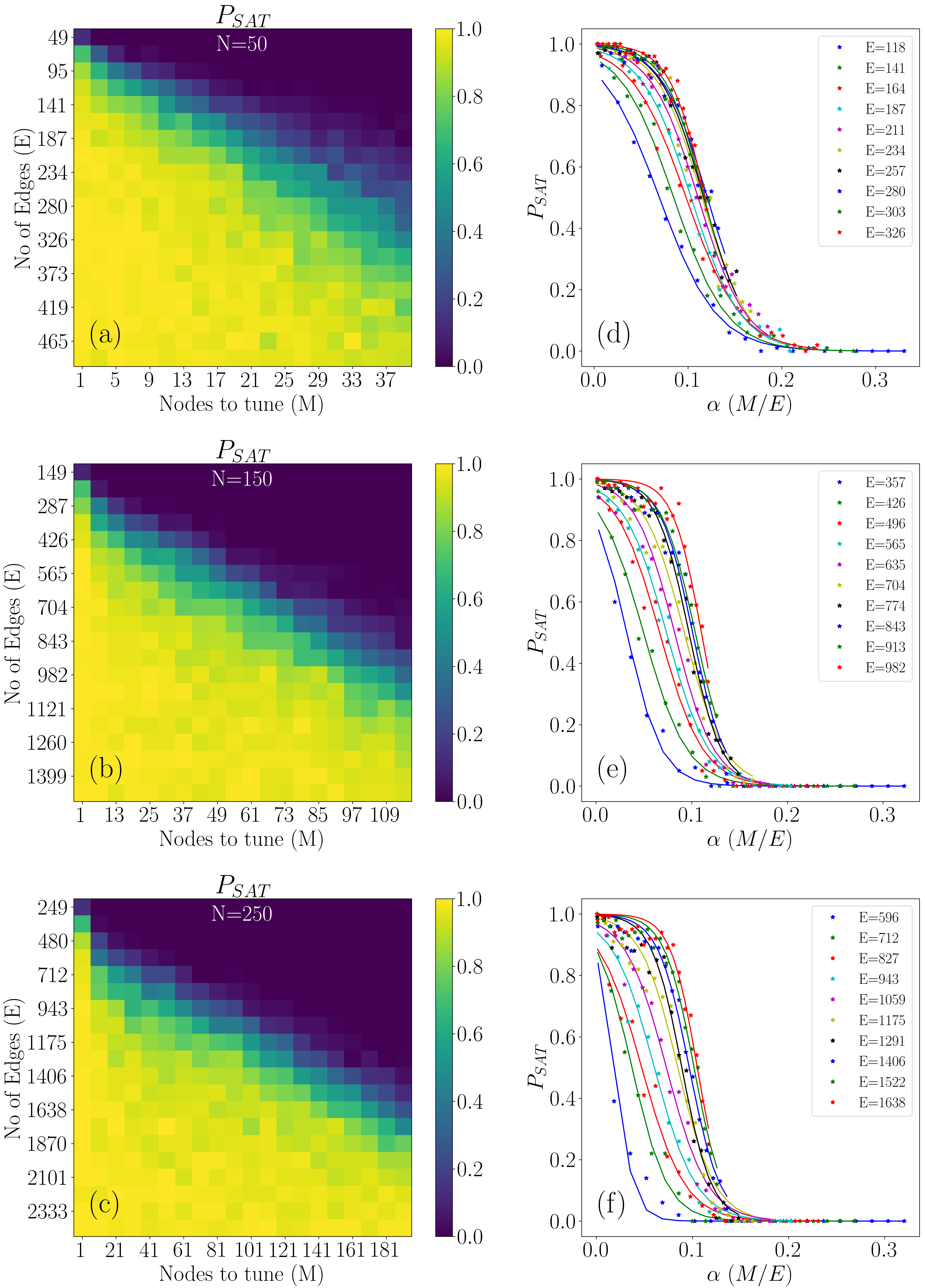}
    \caption{{\it Adaptation phase diagram.}
(a-c) Depictions of $P_{SAT}$ for networks comprising 50, 150, and 250 nodes, respectively. The color bar indicates $P_{SAT}$ values. The x-axis represents the number of output nodes (M), while the y-axis denotes the number of edges (E). 
(d-f) Illustrations of the decline in $P_{SAT}$ in relation to an increase in $\alpha=M/E$. Each curve is representative of a horizontal slice of the 2D phase diagram (displayed on the left) at a constant $E$ value. The data is fitted to a sigmoid like curve as outlined in Eq. \ref{eq:fitting_curve}.}   
    \label{fig:2}
\end{figure}

\begin{figure}[h]
    \centering
    \includegraphics[width=8cm]{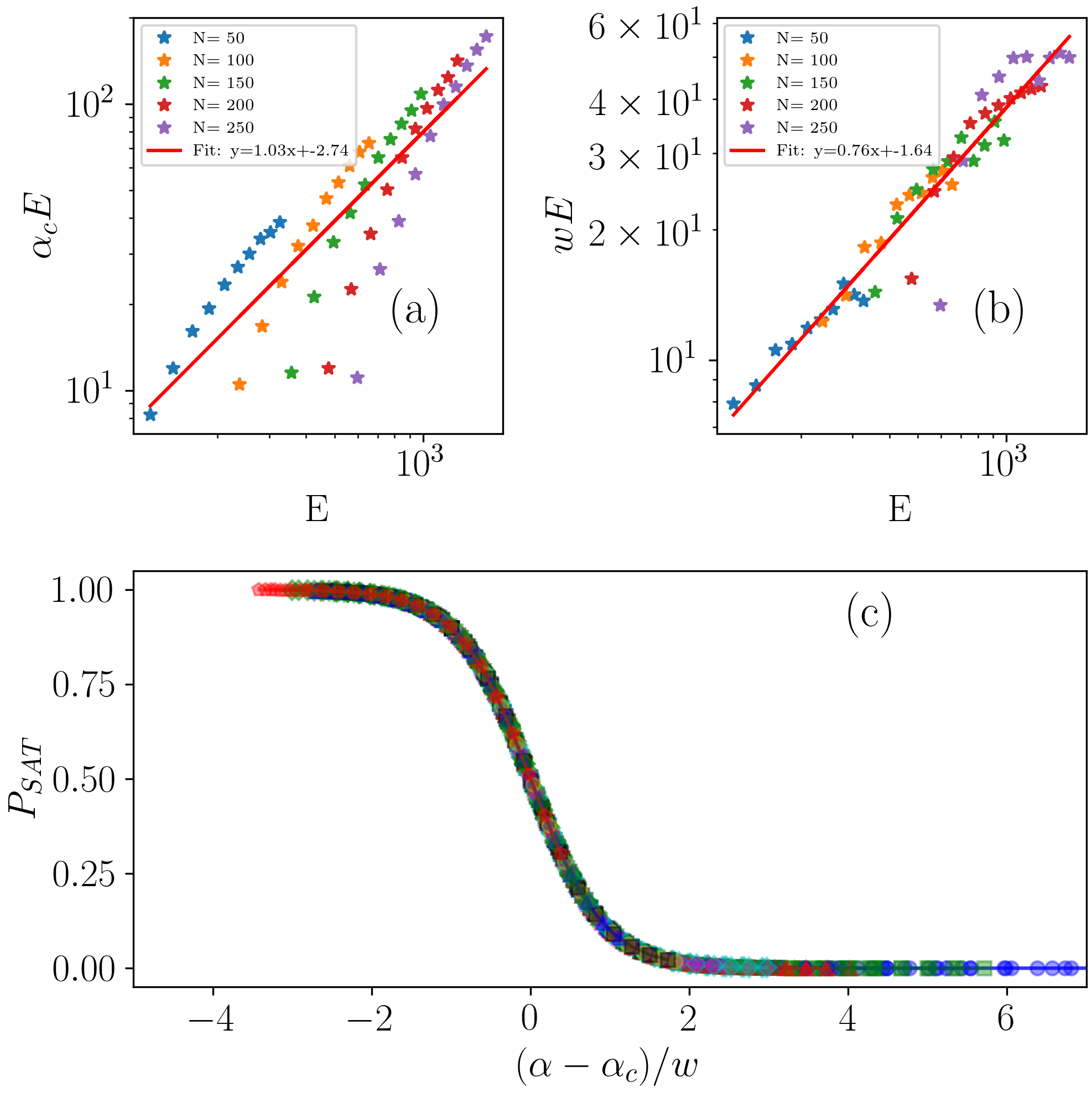}
    \caption{{\it Universal scaling near the transition.} a) Illustrates the scaling of the critical number of nodes that can be tuned with the varying number of edges $E$ for different system sizes. (b) Depicts the scaling of the width around this critical number of nodes with respect to the number of edges $E$ for varying system sizes. (c) Shows that the transition plots collapse onto one another upon rescaling for $N=50,100..250.$}   
    \label{fig:3}
\end{figure}

We implemented the aforementioned procedure on a network comprising 50 nodes and 118 edges, as depicted in Fig. \ref{fig:1}. The conductance values for the flow network were uniformly sampled from the range $[10^{-5},1]$. External pressures were applied to the input nodes (highlighted in green), with each node's external pressure being uniformly sampled from $[-10,10]$. The objective was to adjust the resulting pressure at the output nodes (indicated in red) to align with specified target values. These target pressures were determined by setting $\Delta=1$. Chemicals were introduced at these nodes at intervals of $\tau=5$ units, influencing the conductance of downstream pipes. Successful tuning was evident as the error, represented as $||\mathbf{v}-\mathbf{v_{des}}||$, decreased significantly, and the pressure at the output nodes converged to the desired values.
\\ \\
Fig.~\ref{fig:2} presents the $P_{SAT}$ values across varying parameters: number of edges (E), output nodes (M), total nodes (N), and with a fixed $\Delta=0.1$.  (While simulations were conducted for N = 50, 100, 150, 200, and 250, Fig.\ref{fig:2} specifically displays results for N=50, 150, and 250). The other training parameters remain consistent with those used in Fig.~\ref{fig:1}. We define $P_{SAT}$ as the proportion of networks that achieve successful tuning. A tuning process is deemed successful if the ratio of initial error to final error is less than $10^{-2}$. This ratio was determined for each pixel through 100 training repetitions. A notable surge in $P_{SAT}$ is observed, transitioning from a `hard phase'---where tuning is ineffective---to an `easy phase' where it predominantly succeeds. This phenomenon is further elaborated as a SAT-USAT phase transition in Fig.~\ref{fig:3}. The relevant tuning parameter in such transitions is the clause density $\alpha$~\cite{eli_sat_unsat,monasson2007introduction}, which represents the ratio of clauses to variables. In our context, it is the ratio of nodes to tune to the number of edges ($M/E$). The right column of Fig.~\ref{fig:2} illustrates the decline of $P_{SAT}$ with an increasing $\alpha$. In these plots, curves for constant $E$ are shown. Increasing $\alpha$---achieved by increasing $M$ while maintaining $E$---signifies with an increasing problem hardness. We fit these curves to a sigmoid-like function given by:
\begin{equation}
f(x, a, b) = \frac{1}{2} \left( 1 - \frac{{e^{b \cdot (x - a)} - e^{-b \cdot (x - a)}}}{{e^{b \cdot (x - a)} + e^{-b \cdot (x - a)}}} \right)
\label{eq:fitting_curve}
\end{equation}
From these fits, we deduce the critical point $\alpha_c$ and the transition width $w$. $\alpha_c$ is the $\alpha$ value at which $P_{SAT}=0.5$, and $w$ represents the horizontal span between $P_{SAT}=0.25$ and $P_{SAT}=0.75$.

By analyzing how $\alpha_c$ and $w$ vary with $E$, we deduce the critical number of nodes $M_c=\alpha_c E$ that can be successfully tuned for varying system sizes, specifically for N=50, 100, 150, 200, and 250. The width around this critical value, in terms of the number of nodes for varying system sizes, is given by $\Delta M_c=w E$ (refer to Fig. \ref{fig:3}). We observe that the critical number of nodes $\alpha_c E$ scales as a power law with respect to $E$ with an exponent greater than 1 ($M_c \sim E^{1.03}$). This indicates that our tuning process is scalable and will remain effective for larger system sizes. To demonstrate that this is a SAT-UNSAT phase transition and not just a crossover, we present two supporting arguments: 1) We observe that $w E$ scales with $E$ with an exponent less than 1 ($w E \sim E^{0.76}$). This implies that $w$ scales with $E$ with a negative exponent, indicating that in the thermodynamic limit as $E \rightarrow \infty$, the transition width $w$ vanishes. 2) We note that this transition is universal, as all the transition plots in Fig. \ref{fig:2} (d-f) (and for sizes 100 and 150) collapse onto a universal function upon rescaling by $\alpha \rightarrow (\alpha - \alpha_c)/ w$.

\section{Discussion}
Inspired by Physarum polycephalum, we introduce a physical mechanism to tune flow networks, where each network component uses locally available information to optimise a global cost function. We show that this mechanism is scalable to larger networks and present a phase diagram demonstrating for what values of network parameters successful tuning is observed. Additionally, we showed that this optimization strategy exhibits a SAT-UNSAT phase transition.\\ \\
In previous work, we employed a similar concept to train physical networks~\cite{Anisetti2023}, where the gradient information was encoded in a chemical signal diffused throughout the network. While this approach is equivalent to executing gradient descent, its biological plausibility is questionable due to the slow diffusion-based transport of chemical signals. In contrast, our current methodology involves advective transport of these signals so that the adaptation occurs over faster time scales. However, with this approach, we do not anticipate a gradient descent of the MSE, as the chemical signals only modify the conductances downstream. \\ \\
Given the parallels with particle jamming, which encompasses a SAT-USAT transition~\cite{ruiz-garcia2019tuning}, we propose that an analogous tuning mechanism can be established for hard sphere systems. This has the potential to lay the groundwork for methods to embed memory~\cite{keim2019memory} within these systems and as well learning within these systems given recent work reframing the learning process as feedback-based aging in a glassy landscape~\cite{AnisettiAging}. Moreover, the potential applications of our findings extend further. One immediate and practical application lies in the domain of urban water supply systems. As cities grow and evolve, their water supply networks face the challenge of dynamically adjusting to meet varying demands. The traditional methods of managing these pressures might not be efficient or responsive enough to cater to the rapid changes in demand. Our proposed mechanism offers a potential solution to this challenge, enabling water supply systems to self-optimize and adapt in real-time. Furthermore, the principles elucidated in our study could inspire innovative methodologies for other human-made transport systems. For instance, power grids, which require a delicate balance of supply and demand across vast networks, could benefit from a similar approach. By integrating mechanisms that allow for localized adjustments based on real-time feedback, these systems could become more resilient, efficient, and adaptive.\\ \\
The authors acknowledge Benjamin Scellier and Karen Alim for helpful discussions.  JMS acknowledges financial support from NSF-DMR-2204312. 
\bibliography{biblio.bib} 

\begin{thebibliography}{10}

\bibitem{hu2012blood}
D.~Hu, D.~Cai, and A.~V. Rangan, ``Blood vessel adaptation with fluctuations in
  capillary flow distribution,'' {\em PLOS ONE}, 2012.
\newblock Published: September 27, 2012.

\bibitem{Humphrey2021}
J.~D. Humphrey and M.~A. Schwartz, ``Vascular mechanobiology: homeostasis,
  adaptation, and disease,'' {\em Annual review of biomedical engineering},
  vol.~23, pp.~1--27, 2021.

\bibitem{ronellenfitsch2016global}
H.~Ronellenfitsch and E.~Katifori, ``Global optimization, local adaptation, and
  the role of growth in distribution networks,'' {\em Phys. Rev. Lett.},
  vol.~117, p.~138301, 2016.

\bibitem{Tero2010}
A.~Tero, S.~Takagi, T.~Saigusa, K.~Ito, D.~P. Bebber, M.~D. Fricker, K.~Yumiki,
  R.~Kobayashi, and T.~Nakagaki, ``Rules for biologically inspired adaptive
  network design,'' {\em Science}, vol.~327, no.~5964, pp.~439--442, 2010.

\bibitem{Eytan2003}
D.~Eytan, N.~Brenner, and S.~Marom, ``Selective adaptation in networks of
  cortical neurons,'' {\em Journal of Neuroscience}, vol.~23, no.~28,
  pp.~9349--9356, 2003.

\bibitem{Holland1992}
J.~H. Holland, {\em Adaptation in natural and artificial systems: an
  introductory analysis with applications to biology, control, and artificial
  intelligence}.
\newblock MIT press, 1992.

\bibitem{Mitchell1992}
D.~Mitchell, B.~Selman, H.~Levesque, {\em et~al.}, ``Hard and easy
  distributions of sat problems,'' in {\em Aaai}, vol.~92, pp.~459--465, 1992.

\bibitem{Kirkpatrick1994}
S.~Kirkpatrick and B.~Selman, ``Critical behavior in the satisfiability of
  random boolean expressions,'' {\em Science}, vol.~264, no.~5163,
  pp.~1297--1301, 1994.

\bibitem{Monasson1999}
R.~Monasson, R.~Zecchina, S.~Kirkpatrick, B.~Selman, and L.~Troyansky,
  ``Determining computational complexity from characteristic ‘phase
  transitions’,'' {\em Nature}, vol.~400, no.~6740, pp.~133--137, 1999.

\bibitem{Rocks2019}
J.~W. Rocks, H.~Ronellenfitsch, A.~J. Liu, S.~R. Nagel, and E.~Katifori,
  ``Limits of multifunctionality in tunable networks,'' {\em Proceedings of the
  National Academy of Sciences}, vol.~116, no.~7, pp.~2506--2511, 2019.

\bibitem{Tero2008}
A.~Tero, K.~Yumiki, R.~Kobayashi, T.~Saigusa, and T.~Nakagaki, ``{Flow-network
  adaptation in Physarum amoebae},'' {\em Theory in Biosciences}, vol.~127,
  no.~2, pp.~89--94, 2008.

\bibitem{Alim2017}
K.~Alim, N.~Andrew, A.~Pringle, and M.~P. Brenner, ``{Mechanism of signal
  propagation in Physarum polycephalum},'' {\em Proceedings of the National
  Academy of Sciences of the United States of America}, vol.~114, no.~20,
  pp.~5136--5141, 2017.

\bibitem{monasson2007introduction}
R.~Monasson, ``Introduction to phase transitions in random optimization
  problems,'' {\em arXiv preprint arXiv:0704.2536}, 2007.
\newblock Lectures delivered in Les Houches during the Summer School on Complex
  Systems, July 2006.

\bibitem{Curtis}
E.~Curtis and J.~Morrow, {\em Inverse Problems for Electrical Networks}.
\newblock Series on applied mathematics, World Scientific, 2000.
\newblock pg.40-43, sec. 3.5.

\bibitem{Note1}
For simplicity we assume that the chemical has negligible diffusion and the
  only way it can spread is via the network currents.

\bibitem{Note2}
The current flows into and out of the network through boundary nodes.

\bibitem{Note3}
This helps the potentials at the output node to converge nicely at the desired
  potentials. If this delay is not introduced, the output potential oscillates
  about the desired potential. Also, this helps in avoiding the buildup of
  excess chemical in the network.

\bibitem{eli_sat_unsat}
E.~Chertkov, {\em Phase transitions in random satisfiability problems}.
\newblock 2017.
\newblock Lecture notes available at :
  \url{https://guava.physics.uiuc.edu/~nigel/courses/563/Essays_2017/PDF/chertkov.pdf}.

\bibitem{Anisetti2023}
V.~R. Anisetti, B.~Scellier, and J.~M. Schwarz, ``Learning by non-interfering
  feedback chemical signaling in physical networks,'' {\em Phys. Rev.
  Research}, vol.~5, p.~023024, 2023.

\bibitem{ruiz-garcia2019tuning}
M.~Ruiz-Garc{\'i}a, A.~J. Liu, and E.~Katifori, ``Tuning and jamming reduced to
  their minima,'' {\em Phys. Rev. E}, vol.~100, p.~052608, 2019.

\bibitem{keim2019memory}
N.~C. Keim, J.~D. Paulsen, Z.~Zeravcic, S.~Sastry, and S.~R. Nagel, ``Memory
  formation in matter,'' {\em Rev. Mod. Phys.}, vol.~91, p.~035002, 2019.

\bibitem{AnisettiAging}
V.~R. Anisetti, A.~Kandala, and J.~Schwarz, ``Emergent learning in physical
  systems as feedback-based aging in a glassy landscape,'' {\em arXiv preprint
  arXiv:2309.04382}, 2023.

\end{thebibliography}
\bibliographystyle{ieeetr}
\end{document}